\documentstyle[prl,aps,epsfig,floats]{revtex}

\newcommand\beq{\begin{equation}}
\newcommand\eeq{\end{equation}}
\newcommand\bea{\begin{eqnarray}}
\newcommand\eea{\end{eqnarray}}
\newcommand\non{\nonumber}
\newcommand\bib{\bibitem}

\begin{document}

\draft

\textheight=24cm
\twocolumn[\hsize\textwidth\columnwidth\hsize\csname@twocolumnfalse\endcsname

\title{\Large Gapless Line for the Anisotropic Heisenberg Spin-1/2 Chain in a 
Magnetic Field and the Quantum ANNNI Chain}
\author{Amit Dutta$^{1*}$ and Diptiman Sen$^2$}
\address{\it $^1$ Institut f\"ur Theoretische Physik und Astrophysik, 
Universit\"at W\"urzburg, Am Hubland, 97074 W\"urzburg, Germany \\
$^2$ Centre for Theoretical Studies, Indian Institute of Science, 
Bangalore 560012, India}

\date{\today}
\maketitle

\begin{abstract}
We study the anisotropic Heisenberg ($XYZ$) spin-1/2 chain placed in a 
magnetic field pointing along
the $x$-axis. We use bosonization and a renormalization group analysis to
show that the model has a non-trivial fixed point at a certain value of the 
$XY$ anisotropy $a$ and the magnetic field $h$. Hence, there is a line of 
critical points in the $(a,h)$ plane on which the system is gapless, even 
though the Hamiltonian has no continuous symmetry. The quantum critical line 
corresponds to a spin-flop transition; it separates two gapped phases in one of
which the $Z_2$ symmetry of the Hamiltonian is broken. Our study has a bearing
on one of the transitions of the axial next-nearest neighbor Ising (ANNNI) 
chain in a transverse magnetic field. We also discuss the properties of the 
model when the magnetic field is increased further, in particular, the disorder
line on which the ground state is a direct product of single spin states. 
\end{abstract}
\vskip .5 true cm

\pacs{~~ PACS number: ~75.10.Jm, ~64.70.Rh}
\vskip.5pc
]
\vskip .5 true cm

\section{Introduction}

One-dimensional quantum spin systems have been studied extensively ever since
the problem of the isotropic Heisenberg
spin-1/2 chain was solved exactly by Bethe. Baxter later 
used the Bethe ansatz to solve the anisotropic Heisenberg ($XYZ$) spin-1/2
chain in the {\it absence} of a magnetic field \cite{baxter}; the problem
has not been analytically solved in the presence of a magnetic field.
Experimentally, quantum spin chains and ladders are known to exhibit a wide
range of unusual properties, including both gapless phases with a power-law
decay of the two-spin correlations and gapped phases with an exponential decay
\cite{gogolin,dagotto}. There are also two-dimensional classical statistical
mechanics systems (such as the axial next-nearest neighbor Ising (ANNNI)
model) whose finite temperature properties can be understood by studying an
equivalent quantum spin-1/2 chain
in a magnetic field. The ANNNI model has been studied by several techniques,
and it was believed for a long time to have a floating phase of finite width
in which the system is gapless \cite{villain}. 

Amongst the powerful analytical methods now available for studying quantum 
spin-1/2 chains is the technique of bosonization \cite{gogolin,rao}. Recently,
the $XXZ$ chain in a transverse magnetic field \cite{dmitriev} and the quantum
ANNNI model \cite{allen} have been studied using bosonization. In this paper,
we will study the anisotropic $XYZ$ model in a magnetic field pointing
along the $x$-axis. For small values of the $XY$ anisotropy $a$ and the
magnetic field $h$, we will show in Sec. II that there is a non-trivial fixed 
point (FP) of the renormalization group (RG) in the $(a,h)$ plane; the system 
is gapless on a quantum critical line of points which flow to this FP. In Sec.
III, one of the transitions of the ANNNI model will be shown to be a 
special case of our results in which the $zz$ coupling is equal to zero. 
Our results are complimentary to the earlier studies of the ANNNI model which 
indicate a gapless phase of finite width. We will present the complete zero 
temperature phase diagram of the ANNNI model which has both a gapless phase 
of finite width as well as a gapless line.

The gapless line is somewhat unusual because the $XY$ anisotropy
and the magnetic field both break the continuous symmetry of rotations in the 
$x-y$ plane. In Sec. IV, we will provide a physical understanding of the
gapless line by going to the classical (large $S$) limit of the model; this 
helps us to identify it as a spin-flop transition line. In Sec. V, we will 
discuss a disorder line which lies at a larger value of the magnetic field.
In Sec. VI, we will briefly comment on the Ising transition which
occurs at an even larger value of the magnetic field.

\section{Bosonization and Renormalization Group Analysis}

We consider the anisotropic Hamiltonian defined on a chain of sites
\bea
H ~=~ \sum_n ~[~ & & (1+a) ~S_n^x S_{n+1}^x ~+~ (1-a)~ S_n^y S_{n+1}^y \non \\
& & +~ \Delta ~S_n^z S_{n+1}^z ~-~ h ~S_n^x ~]~,
\label{ham1}
\eea
where the $S_n^{\alpha}$ are spin-1/2 operators.
We will assume that the $XY$ anisotropy $a$ and the $zz$ coupling $\Delta$
satisfy $-1 \le a, \Delta \le 1$. We can assume without loss of generality 
that the magnitude of the $zz$ coupling is smaller than the $yy$ coupling 
(i.e., $|\Delta| < 1 - a$), and that the magnetic field strength $h \ge 0$.
The Hamiltonian in Eq. (\ref{ham1}) is invariant under the global $Z_2$ 
transformation $S_n^x \rightarrow S_n^x, S_n^y \rightarrow - S_n^y, S_n^z 
\rightarrow - S_n^z$.

For $a=h=0$, the model is symmetric under rotations in the $x-y$ plane and is 
gapless. The low-energy and long-wavelength modes of the system are then 
described by the bosonic Hamiltonian \cite{gogolin,rao}
\beq
H_0 ~=~ \frac{v}{2} ~\int ~dx ~[~ (\partial_x \theta)^2 ~+~ (\partial_x
\phi)^2 ~]~,
\label{ham2}
\eeq
where $v$ is the velocity of the low-energy excitations (which have the
dispersion $\omega = v |k|$); $v$ is a function of $\Delta$.
(The continuous space variable $x$ and the site label $n$ are related through
$x=nd$, where $d$ is the lattice spacing.) The bosonic theory contains another
parameter called $K$ which is related to $\Delta$ by \cite{gogolin,rao}
\beq
K ~=~ \frac{\pi}{\pi + 2 \sin^{-1} (\Delta)} ~.
\label{kd}
\eeq
$K$ takes the values 1 and 1/2 for $\Delta =0$ (which describes noninteracting
spinless fermions) and $\Delta =1$ (the isotropic antiferromagnet)
respectively; as $\Delta \rightarrow -1$, $K \rightarrow \infty$. We thus
have $1/2 \le K < \infty$.

In terms of the fields $\phi$ and $\theta$ introduced in Eq. (\ref{ham2}),
the spin operators can be written as \cite{dmitriev}
\bea
S_n^z ~&=&~ {\sqrt {\frac{\pi}{K}}} ~\partial_x \phi ~+~ (-1)^n c_1
\cos (2 {\sqrt {\pi K}} \phi ) ~, \non \\
S_n^x ~&=&~ [~ c_2 \cos (2 {\sqrt {\pi K}} \phi) ~+~ (-1)^n c_3 ~]~
\cos ({\sqrt {\frac{\pi}{K}}} \theta ) ~,
\label{spin}
\eea
where the $c_i$ are constants given in Ref. 8. The $XY$ anisotropy term is
given by
\beq
S_n^x S_{n+1}^x ~-~ S_n^y S_{n+1}^y ~=~ c_4 \cos (2{\sqrt {\frac{\pi}{K}}}
\theta ) ~,
\label{anis}
\eeq
where $c_4$ is another constant.

For convenience, let us define the three operators
\bea
{\cal O}_1 ~&=&~ \cos (2 {\sqrt {\pi K}} \phi) \cos ({\sqrt {\frac{\pi}{K}}}
\theta ) ~, \non \\
{\cal O}_2 ~&=&~ \cos (2 {\sqrt {\frac{\pi}{K}}} \theta ) ~, \quad {\rm and}
\quad {\cal O}_3 ~=~ \cos (4 {\sqrt {\pi K}} \phi) ~.
\label{ops}
\eea
Their scaling dimensions are given by $K + 1/4K$, $1/K$ and $4K$ respectively.
Using Eqs. (\ref{spin}-\ref{anis}), the terms corresponding to $a$ and $h$ in
Eq. (\ref{ham1}) can be written as
\beq
H_a + H_h ~=~ \int dx ~[~ a c_4 {\cal O}_2 - h c_2 {\cal O}_1 ~]~ ,
\eeq
where we have dropped rapidly varying terms proportional to $(-1)^n$ since
they will average to zero in the continuum limit. (We will henceforth absorb
the factors $c_4$ ($c_2$) in the definitions of $a$ ($h$).)
We will now study how the parameters $a$ and $h$ flow under RG.

The operators in Eqs. (\ref{ops}) are related to each other through the
operator product expansion; the RG equations for their coefficients
will therefore be coupled to each other \cite{cardy}. In our model,
this can be derived as follows. Given
two operators $A_1 = \exp (i\alpha_1 \phi + i\beta_1 \theta)$ and $A_2 =
\exp (i\alpha_2 \phi + i\beta_2 \theta)$, we write the fields $\phi$ and
$\theta$ as the sum of slow fields (with wave numbers $|k| < \Lambda e^{-dl}$)
and fast fields (with wave numbers $\Lambda e^{-dl} < |k| < \Lambda$), where
$\Lambda$ is the momentum cut-off of the theory, and $dl$ is the change in
the logarithm of the length scale. Integrating out the fast fields shows that
the product of $A_1$ and $A_2$ at the same space-time point gives a
third operator $A_3 = e^{i(\alpha_1 + \alpha_2 )\phi + i(\beta_1 + \beta_2 )
\theta}$ with a prefactor which can be schematically written as
\beq
A_1 A_2 ~\sim ~ e^{-(\alpha_1 \alpha_2 + \beta_1 \beta_2 ) dl /2\pi} ~A_3 ~.
\eeq
If $\lambda_i (l)$ denote the coefficients of the operators $A_i$ in an
effective Hamiltonian, then the RG expression for $d\lambda_3 /dl$ will contain
the term $(\alpha_1 \alpha_2 + \beta_1 \beta_2 ) \lambda_1 \lambda_2 /2\pi$.
Using this, we find that if the three operators in Eqs. (\ref{ops}) have
coefficients $h$, $a$ and $b$ respectively, then the RG equations are
\bea
\frac{dh}{dl} ~&=&~ (2 - K - \frac{1}{4K}) h ~-~ \frac{1}{K} ah ~-~ 4Kbh ~,
\non \\
\frac{da}{dl} ~&=&~ (2 -\frac{1}{K}) a ~-~ (2K -\frac{1}{2K}) h^2 ~, \non \\
\frac{db}{dl} ~&=&~ (2 - 4K) b ~+~ (2K -\frac{1}{2K}) h^2 ~, \non \\
\frac{dK}{dl} ~&=&~ \frac{a^2}{4} ~-~ K^2 b^2 ~,
\label{rg}
\eea
where we have absorbed some factors involving $v$ in the variables $a$, $b$
and $h$. (We will ignore the RG equation for $v$ here.) It will turn out that
$K$ renormalizes very little in the regime of RG flows that we will be
concerned with. Eqs. (\ref{rg}) have appeared earlier in the context of some
other problems \cite{nersesyan,giamarchi}. However, the last two
terms in the expression for $dh/dl$ were not presented in Ref. 10; these two
terms turn out to be crucial for what follows. Note that Eqs. (\ref{rg}) are
invariant under the duality transformation $K \leftrightarrow 1/4K$ and
$a \leftrightarrow b$.

Let us now consider the fixed points of Eqs. (\ref{rg}). For any value of
$K=K^*$, a trivial FP is $(a^* ,b^* ,h^* )=(0,0,0)$. Remarkably, it turns out
that there is a non-trivial FP for any value of $K^*$ lying in
the range $1/2 < K^* < 1 + {\sqrt 3}/2$; we will henceforth restrict our
attention to this range of values. (The upper bound on $K^*$ comes from the
condition $2-K^*-1/4K^* > 0$.) The non-trivial FP is given by
\bea
h^* ~&=&~ \frac{\sqrt {2K^*(2-K^*-1/4K^*)}}{2K^*+1} ~, \non \\
a^* ~&=&~ (K^*+\frac{1}{2}) h^{*2} ~, \quad {\rm and} \quad b^* ~=~
\frac{a^*}{2K^*} ~.
\label{fp}
\eea
The system is gapless at this FP as well as at all points which flow to this
FP. One might object that Eqs. (\ref{rg}) can only be trusted if $a$, $b$
and $h$ are not too large, otherwise one should go to higher orders.
We note that the FP approaches the origin as $K^* \rightarrow 1
+ {\sqrt 3}/2 \simeq 1.866$; from Eq. (\ref{kd}), 
this corresponds to the $zz$ coupling $\Delta
= - \sin [\pi ({\sqrt 3} - 3/2)] \simeq - 0.666$. Thus the RG equations
can certainly be trusted for $K^*$ close to 1.866. For $K^*=1$, the FP is
at $(a^* , b^* , h^*) = (1/4,1/8, 1/{\sqrt 6})$.

We have numerically studied the RG flows given by Eqs. (\ref{rg}) for various
starting values of $(K,a,b,h)$. Since the Hamiltonian in
Eq. (\ref{ham1}) does not contain the operator ${\cal O}_3$, we set $b=0$
initially. We take $a$ and $h$ to be very small initially, and see which
set of values flows to a non-trivial FP. For instance, starting with
$K=1$, $b=0$, and $a,h$ very small, we find that there is a line of points
which flow to a FP at $(K^* , a^*, b^*, h^*) = (1.020, 0.246, 0.122, 0.404)$.
This line projected on to the $(a,h)$ plane is shown in Fig. 1. We see
that $K$ changes very little during this flow; if we start with a larger value
of $K$ initially, then it changes even less as we go to the non-trivial
FP. It is therefore not a bad approximation to ignore the flow of
$K$ completely.

\begin{figure}[htb]
\begin{center}
\hspace*{-0.5cm}
\epsfig{figure=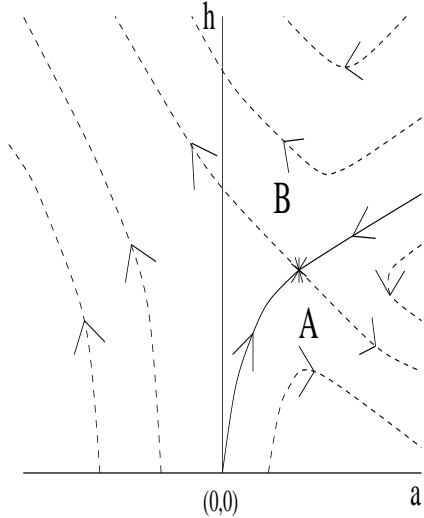,width=7.5cm,height=6cm,angle=270}
\end{center}
\caption{RG flow diagram in the $(a,h)$ plane. The solid line shows the set of
points which flow to the FP at $a^*=0.246$, $h^*=0.404$ marked by an asterisk.
The dotted lines show the RG flows in the gapped phases A and B (see text).}
\end{figure}

We can characterize the set of points $(a,h)$ lying close to the
origin which flow to the non-trivial FP. Numerically, we find that there is
a unique flow line in the $(a,h)$ plane for each starting value of $K$ and
$b=0$, provided that $a, h$ are very small initially. This means
that $a(l)$ and $h(l)$ given by Eqs. (\ref{rg}) must follow the same line
regardless of the starting values of $a,h$. From Eqs. (\ref{rg}), we see that
if $h << a^{1/2}$, then $h(l) \sim h(0) \exp (2-K-1/4K)l$ while $a(l) \sim
\exp (2-1/K)l$. Hence $h$ must initially scale with $a$ as
\beq
h ~\sim ~a^{(2-K-1/4K)/(2-1/K)} ~,
\label{scaling}
\eeq
as we have numerically verified for $K=1$.
However, Eq. (\ref{scaling}) is only true if $(2-K-1/4K)/(2-1/K) > 1/2$, i.e.,
if $K < (1 + {\sqrt 2})/2 =1.207$. For $K \ge 1.207$ (i.e., $\Delta \le
-0.266$), the initial scaling form is given by $h \sim a^{1/2}$.

We now examine the stability of small perturbations
away from the fixed points. The trivial FP
at the origin has two unstable directions ($a$ and $h$),
one stable direction ($b$) and one marginal direction ($K$). The non-trivial
FP has two stable directions, one unstable direction and a marginal
direction (which corresponds to changing $K^*$ and simultaneously $a^*$, $b^*$
and $h^*$ to maintain the relations in Eqs. (\ref{fp})). The presence of two
stable directions implies that there is a two-dimensional surface of
points (in the space of parameters $(a,b,h)$) which flows to this FP;
the system is gapless on that surface. A perturbation in the unstable
direction produces a gap in the spectrum. For
instance, at the FP with $(K^* ,a^* ,b^* ,h^*)=(1,1/4,1/8, 1/{\sqrt 6})$,
the four RG eigenvalues are given by 1.273 (unstable), 0 (marginal), and
$-1.152 \pm 1.067 i$ (both stable). The positive eigenvalue corresponds to
an unstable direction given by $(\delta K, \delta a, \delta b, \delta h)=
\delta a (0.113, 1, -0.092, -0.239)$. A small perturbation of size $\delta a$
in that direction will produce a gap in the spectrum which scales as
$\Delta E \sim |\delta a|^{1/1.273} = |\delta a|^{0.786}$; the correlation
length is then given by $\xi \sim v/\Delta E \sim |\delta a|^{-0.786}$.

Figure 1 shows that the set of points which do not flow to the non-trivial
FP belong to either region A or region B. These regions can be
reached from the non-trivial FP by moving in the unstable direction, with
$\delta a > 0$ for region A, and $\delta a < 0$ for region B. In region A,
the points flow to $a=\infty$; this corresponds to a gapped phase in which
the the $xx$ coupling is larger than the $yy$ and $zz$ couplings. In region 
B, both $a$ and $h$ flow to $-\infty$; this is a gapped phase in which the 
$yy$ coupling is larger than the $xx$ and $zz$ couplings. We will now see that
the difference between these two phases lies in the way in which the $Z_2$ 
symmetry of the Hamiltonian is realized. An order parameter which 
distinguishes between the two phases is the staggered magnetization in the 
$y$ direction, defined in terms of a ground state expectation value as
\beq
m_y ~=~ [~ \lim_{n \rightarrow \infty} ~(-1)^n < S_0^y S_n^y > ~]^{1/2} ~.
\label{order}
\eeq
This is zero in phase A; hence the $Z_2$ symmetry is unbroken. In phase B, 
$m_y$ is non-zero, and the $Z_2$ symmetry is broken. The scaling of $m_y$ 
with the perturbation $\delta a$ can be found
as follows \cite{dmitriev}. At $a=h=0$, the leading term in the long-distance
equal-time correlation function of $S^y$ is given by
\beq
< S_0^y ~S_n^y > ~\sim ~ \frac{(-1)^n}{|n|^{1/2K}} ~.
\eeq
Hence the scaling dimension of $S_n^y$ is $1/4K$. In a gapped
phase in which the correlation length is much larger than
the lattice spacing, $m_y$ will therefore scale with the gap as $m_y \sim
(\Delta E)^{1/4K}$. If we assume that the scaling dimension of $S_n^y$ at the
non-trivial FP remains close to $1/4K$, then the numerical
result quoted in the previous paragraph for $K=1$ implies that $m_y \sim
|\delta a|^{0.196}$ for $\delta a$ small and negative.

The nature of the transition on the gapless line will be discussed in Sec. IV.
We will argue there that this is a spin-flop transition line. (Spin-flop 
transitions in one-dimensional spin-1/2 chains have been studied earlier 
\cite{karadamoglou,kenzelmann,sakai}.) 

\section{Quantum ANNNI Model}

We will now apply our results to the one-dimensional spin-1/2 quantum ANNNI
model \cite{villain,allen}, with nearest neighbor ferromagnetic and 
next-nearest neighbor antiferromagnetic Ising interactions and a transverse 
magnetic field. The Hamiltonian is given by
\beq
H_A ~=~ \sum_n ~[~ - 2 J_1 T_n^x T_{n+1}^x + J_2 T_n^x T_{n+2}^x +
\frac{\Gamma}{2} T_n^y ~]~,
\label{annni1}
\eeq
where $J_1 , J_2 >0$, and the $T_n^{\alpha}$ are spin-1/2 operators; we 
can assume without loss of generality that
$\Gamma \ge 0$. The quantum Hamiltonian in Eq. (\ref{annni1}) is related
to the transfer matrix of the two-dimensional classical ANNNI model; the
finite temperature critical points of the latter are related to the ground 
state quantum critical points of Eq. (\ref{annni1}), with the temperature
$T$ being related to the magnetic field $\Gamma$.

Some earlier studies showed that the model has a floating phase of finite width
which is gapless \cite{villain}. A recent bosonization study reached the same 
conclusions \cite{allen}. (Recent numerical studies of the two-dimensional 
classical ANNNI model at finite temperature have led to contradictory results 
for the width of the floating phase \cite{shirahata}.) All these
studies (both analytical and numerical) indicate that the phase transition is 
of the Kosterlitz-Thouless type (with $\xi$ diverging exponentially) from the
high-temperature side (i.e., from region B in Fig. 1 for the quantum ANNNI
model), and is of the Pokrovsky-Talapov type \cite{dzhaparidze} (with $\xi$ 
diverging as a power-law) from the low-temperature side (i.e., from region 
A in Fig. 1). 

We will now apply our results to the quantum ANNNI model.
Consider a Hamiltonian which is dual to Eq. (\ref{annni1}) for spin-1/2;
this will turn out to be a special case of our earlier model. The dual
Hamiltonian is given by \cite{villain,sen}
\beq
H_D ~=~\sum_n ~[~J_2 S_n^x S_{n+1}^x ~+~\Gamma S_n^y S_{n+1}^y -J_1 S_n^x ~]~,
\label{annni2}
\eeq
where $S_n^{\alpha}$ are the spin-1/2 operators dual to $T_n^{\alpha}$
(for instance, $S_n^x = 2 T_n^x T_{n+1}^x$ and $T_n^y = 2 S_{n-1}^y S_n^y$). 
After scaling this Hamiltonian by an appropriate factor, we see that it
has the same form as in Eq. (\ref{ham1}), with 
\bea
a ~&=&~ \frac{J_2 - \Gamma}{J_2 + \Gamma} ~, \non \\
h ~&=&~ \frac{2J_1}{J_2 + \Gamma} ~, \non \\
\Delta ~&=&~ 0 ~.
\eea
Hence it follows that the quantum ANNNI model has a line of points in the 
$(J_2/J_1, \Gamma /J_1)$ plane on which the system is gapless. From Eq. 
(\ref{scaling}), we see that the shape of this line is given by $J_1 \sim 
(J_2 - \Gamma)^{3/4}$ as $J_1 \rightarrow 0$.

The analysis in Sec. II indicates that as the transition line is approached,
$\xi$ should diverge as a power-law from both sides. We now have to 
reconcile this with some of the earlier 
analytical \cite{villain,allen} and numerical \cite{shirahata} studies 
which showed that as one approaches the floating phase, $\xi$ diverges as a 
power-law from phase B but exponentially from phase A. The important point is 
that these earlier studies were carried out at values of $J_2 /J_1$ which are 
close to 1, while our RG results are expected to be valid 
only if $a,h$ are small, i.e., if $J_2 /J_1$ is large.
If $J_2/J_1$ is close to 1, the situation is quite different for the following 
reason. Exactly at $J_2/J_1 =1$ and $\Gamma =0$, the Hamiltonian in 
Eq. (\ref{annni2}) can be written in the form 
\beq
H_{MC} ~=~ J_2 ~\sum_n ~(S_n^x ~-~ \frac{1}{2})~(S_{n+1}^x ~-~ \frac{1}{2})~.
\eeq
This is a multicritical point with a ground state degeneracy growing 
exponentially with the system size, since any state in which every pair
of neighboring sites $(n,n+1)$ has at least one site with $S^x =1/2$ is a
ground state. We can now study what happens when we go slightly away from this
multicritical point. To lowest order, this involves doing perturbation theory 
within the large space of degenerate states. An argument due to Villain 
and Bak \cite{villain} shows that 
if $J_2 - J_1$ and $\Gamma$ are non-zero but small, then the low-energy 
properties of Eq. (\ref{annni2}) do not change if $\Gamma S_n^y S_{n+1}^y$ is
replaced by $(\Gamma /2) (S_n^y S_{n+1}^y + S_n^z S_{n+1}^z)$. (This is 
because the difference between the two kinds of terms is given by operators 
which, acting on one of the degenerate ground states, take it out of the 
degenerate space to a higher excited 
state in which a pair of neighboring sites have $S^x = -1/2$.)
Thus the fully anisotropic model becomes equivalent to a different model
which is invariant under the $U(1)$ symmetry of rotations in the $y-z$ plane.
The $U(1)$ symmetric model has been studied earlier using bosonization
\cite{gogolin,giamarchi,cabra}; it has a gapless phase of finite width which
lies between two gapped phases. Thus the difference between our study (in 
which $J_2 - \Gamma$ and $J_1$ are small) and the earlier studies (in which 
$J_2 - J_1$ and $\Gamma$ are small) is that they have different symmetries 
away from the transition line, namely, $Z_2$ and $U(1)$ respectively. Our 
study and the earlier studies are therefore complimentary to each other; a 
combination of the two leads to a complete understanding of the model over
the entire parameter range.

To summarize, the transition from phase A to phase B can occur either 
through a gapless line (if $a,h$ are small), or through a gapless phase of
finite width (if $a,h$ are large). The complete phase diagram of the ground 
state of Eq. (\ref{annni2}) is shown in Fig. 2 \cite{villain}. The three major
phases shown are distinguished by the following properties of the expectation 
values of the different components of the spins. In the antiferromagnetic 
phase, the spins point alternately along the ${\hat x}$ and $-{\hat x}$
directions. In the spin-flop phase, they point alternately along the ${\hat y}$
and $-{\hat y}$ directions, with an uniform tilt towards the ${\hat x}$ 
direction. In the ferromagnetic phase, all the spins point predominantly in 
the ${\hat x}$ direction. The antiferromagnetic and spin-flop
phases are separated by a floating
phase of finite width for $J_2/J_1$ close to 1/2, and by a spin-flop transition
line for large values of $J_2/J_1$. We conjecture that the floating phase and 
the spin-flop transition line are separated by a Lifshitz point as indicated
in Fig. 2. The disorder line and the Ising transition are discussed in Secs. 
V and VI respectively. 

We should point out here that in terms of the original Hamiltonian in Eq. 
(\ref{annni1}), some of the phases shown in Fig. 2 have somewhat different 
names \cite{villain}. The spin-flop phase is called the paramagnetic phase; 
this is further divided into two phases by the disorder line, namely, a 
commensurate phase to the left and an incommensurate phase on the right of 
the disorder line. The antiferromagnetic phase is called the antiphase. 

\begin{figure}[htb]
\begin{center}
\hspace*{-0.5cm}
\epsfig{figure=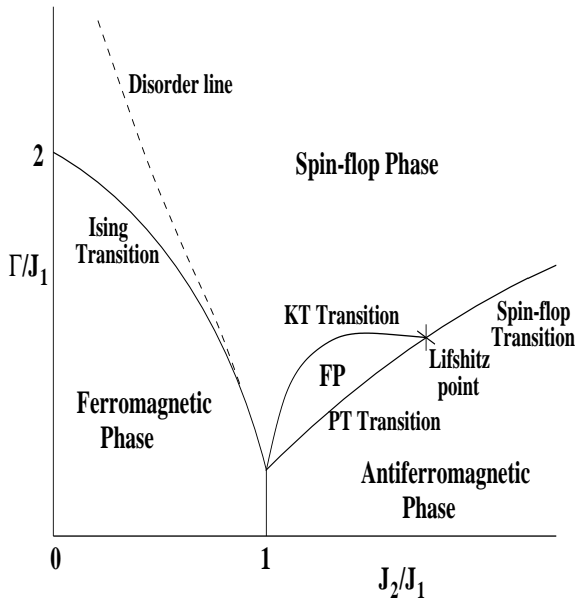,width=9cm,height=8cm,angle=270}
\end{center}
\caption{Schematic phase diagram of the model described in Eq. (\ref{annni2}).
The various phases and transition lines are explained in the text. The 
initials FP, KT and PT stand for floating phase, Kosterlitz-Thouless and
Pokrovsky-Talapov respectively.} 
\end{figure}

\section{Classical Limit}

In this section, we would like to provide a physical picture of the gapless 
line in the $(a,h)$ plane by looking at the classical limit of Eq. 
(\ref{ham1}). Consider the Hamiltonian
\bea
H_{S1} ~=~ \sum_n ~[~ & & (1+a) ~S_n^x S_{n+1}^x ~+~ (1-a) ~S_n^y S_{n+1}^y
\non \\
& & +~ \Delta ~S_n^z S_{n+1}^z ~-~ 2 S h ~S_n^x ~]~,
\label{ham3}
\eea
where the spins satisfy ${\bf S}_n^2 = S(S+1)$, and we are interested in the
classical limit $S \rightarrow \infty$ \cite{sen}. (We have multiplied
the magnetic field by a factor of $2S$ in Eq. (\ref{ham3}) so that we recover 
Eq. (\ref{ham1}) for spin-1/2.) We assume as before that
the $zz$ coupling is smaller in magnitude than the $yy$ coupling.
Then the classical ground state of Eq. (\ref{ham3})
is given by a configuration in which all the spins lie in the $x-y$ plane,
with the spins on odd and even numbered sites pointing respectively at an
angle of $\alpha_1$ and $-\alpha_2$ with respect to the $x$-axis.
The ground state energy per site is
\bea
e (\alpha_1 , \alpha_2 ) ~=~ S^2 ~[ & & - ~h ~(\cos \alpha_1 +
\cos \alpha_2 ) ~+~ \cos (\alpha_1 + \alpha_2 ) \non \\
& & + ~a ~\cos (\alpha_1 - \alpha_2 ) ].
\label{solns}
\eea
Minimizing this with respect to $\alpha_1$ and $\alpha_2$, we discover that
there is a special line given by $h^2 ~=~ 4 a$ on which {\it all} solutions
of the equation
\beq
h ~\cos (\frac{\alpha_1 - \alpha_2}{2} ) ~=~ 2 ~\cos (\frac{\alpha_1 +
\alpha_2}{2} ) ~,
\label{gnst}
\eeq
give the same ground state energy per site, $e_0 =-(1+a)S^2$. The solutions of
Eq. (\ref{gnst}) range from $\alpha_1 = \alpha_2 = \cos^{-1} (h/2)$ to
$\alpha_1 = \pi, \alpha_2 = 0$ (or vice versa); in the ground state phase
diagram of the ANNNI model, these two configurations correspond respectively
to a antiferromagnetic alignment of the spins with respect to the $y$-axis
(with a small tilt towards the $x$-axis if $h$ is small), and an
antiferromagnetic alignment of the spins with respect to the $x$-axis. 
The curve $h^2 = 4a$ is therefore a phase transition line, and the form of 
the ground states on the two sides shows that there is a spin-flop transition 
across that line. Further, we see that for $h^2 =4a$, there is a one-parameter 
set of classical ground states (characterized by, say, the value of 
$\alpha_1$ which can go all the way from 0 to $2\pi$ in the solutions of 
Eq. (\ref{gnst})) which are all degenerate. Hence the symmetry is enhanced
from a $Z_2$ symmetry away from the line to a $U(1)$ symmetry (of rotations 
in the $x-y$ plane) on the line. We therefore expect a gapless mode in the 
excitation spectrum corresponding to the Goldstone mode of the broken 
continuous symmetry. We can find this gapless 
mode explicitly by going to the next order in a $1/S$ expansion \cite{sen}. 

The above arguments provide some understanding of why one may expect such a 
gapless line in the spin-1/2 model also. Note however that the bosonization 
analysis gives the scaling form in Eq. (\ref{scaling}) for $h$ versus $a$; this
agrees with the classical form only if $\Delta \le -0.266$. Further, in the 
classical limit, the transition across the gapless line is of first order, 
whereas it is of second order in the spin-1/2 case. There is probably a
critical value of the spin $S$ above which the transition is of first order.
(For the $U(1)$ symmetric model described by Eq. (\ref{ham4}) below, it is
known that the transition is of first order if $S \ge 1$ \cite{sakai}.)

The classical limit also makes it clear why our model has a different
behavior from the $U(1)$ symmetric model governed by the Hamiltonian
\bea
H_{S2} ~=~ \sum_n ~[~ & & (1+a) ~S_n^x S_{n+1}^x ~+~ (1-a)~S_n^y S_{n+1}^y
\non \\
& & +~ (1-a) ~S_n^z S_{n+1}^z ~-~ 2 S h ~S_n^x ~]~.
\label{ham4}
\eea
In the limit $S \rightarrow \infty$, there is now a {\it two-parameter} set 
of degenerate ground states on the line $h^2 = 4a$; these are obtained by
taking the one-parameter family of configurations given in Eq. (\ref{solns}) 
and rotating them by an arbitrary angle about the $x$-axis. Hence, the 
symmetry of this model is enhanced from $U(1)$ to $SU(2)$ on the line $h^2 =
4a$, and there are now two Goldstone modes instead of one. Considering this 
difference in symmetry for large $S$, it is not surprising that even the 
spin-1/2 models with $U(1)$ symmetry and $Z_2$ symmetry respectively exhibit 
very different behaviors at the spin-flop transition line.

\section{Disorder Line}

We have seen that as the magnetic field $h$ is increased from zero for the 
spin-1/2 model described by Eq. (\ref{ham1}), there is a spin-flop transition 
at a critical field $h_c$ whose value depends on $a$ and $\Delta$. One might 
wonder what happens if the field is increased well beyond $h_c$. 

It turns out that above $h_c$, there is an interesting value of
the field $h=h_d$ where the ground state of the model is exactly solvable 
\cite{peschel,muller}. This field is given by
\beq
h_d ~=~ {\sqrt {2 (1+a+\Delta)}} ~.
\label{hd}
\eeq
At this point, the ground state has a very simple direct product form in
which all the spins lie in the $x-y$ plane, with the spins on even and odd
sublattices pointing at the angles $\alpha$ and $-\alpha$ respectively 
with respect to the $x$-axis, where
\beq
\alpha ~=~ \cos^{-1} (\frac{h_d}{2}) ~.
\label{alpha}
\eeq
To show that this configuration is the ground state of the Hamiltonian, we
observe that the Hamiltonian can be written, up to a constant, as the 
following sum
\bea
H &=& \sum_n ~[H_{2n,2n+1} ~+~ H_{2n,2n-1}]~, \non \\
H_{2n,2n \pm 1} &=& ~~ (\cos \alpha ~S_{2n}^x +
\sin \alpha ~S_{2n}^y - \frac{1}{2}) \non \\
& & \times (\cos \alpha ~S_{2n \pm 1}^x + \sin \alpha ~ 
S_{2n \pm 1}^y - \frac{1}{2}) \non \\
& & + (\cos \alpha ~S_{2n}^x - \sin \alpha ~S_{2n}^y -
\frac{1}{2}) \non \\
& & \times (\cos \alpha ~S_{2n \pm 1}^x - \sin \alpha ~
S_{2n \pm 1}^y - \frac{1}{2}) \non \\
& & +~ \Delta ~[~ \frac{1}{4} ~- (\cos \alpha ~S_{2n}^x + \sin \alpha ~
S_{2n}^y) \non \\ 
& & ~~~~~~~~~~~~ \times (\cos \alpha ~S_{2n \pm 1}^x - \sin \alpha ~
S_{2n \pm 1}^y) \non \\
& & ~~~~~~~~~~~~ - (\sin \alpha ~S_{2n}^x - \cos \alpha ~S_{2n}^y) \non \\
& & ~~~~~~~~~~~~ \times (\sin \alpha ~S_{2n \pm 1}^x + \cos \alpha ~
S_{2n \pm 1}^y) \non \\
& & ~~~~~~~~~~~~ + ~S_{2n}^z S_{2n \pm 1}^z ~] ~,
\label{ham5}
\eea 
where $\alpha$ is given in Eqs. (\ref{hd}-\ref{alpha}). We now use the theorem
that the ground state energy of $H$ is greater than or equal to the sum of the
ground state energies of $H_{2n,2n \pm 1}$, with equality holding if and only 
if there is a state which is simultaneously an eigenstate of all the 
$H_{2n,2n \pm 1}$. Now, each of the Hamiltonians $H_{2n,2n \pm 1}$ in Eq. 
(\ref{ham5}) is a sum of three operators whose eigenvalues are non-negative if
$\Delta \ge 0$ \cite{muller}. The state described in Eq. (\ref{alpha}),
in which all the spins on the even sublattice satisfy $\cos \alpha 
S_{2n}^x + \sin \alpha S_{2n}^y = 1/2$ and all the spins on the odd sublattice
satisfy $\cos \alpha S_{2n+1}^x - \sin \alpha S_{2n+1}^y = 1/2$, is the 
ground state of all the Hamiltonians in Eq. (\ref{ham5}) with zero eigenvalue.
We can actually show, by looking at a two-site system governed by
a single Hamiltonian $H_{2n,2n+1}$, that even if $\Delta < 0$, the state
described above is its ground state provided that $1 - a \ge - \Delta$, i.e., 
as long as the magnitude of the $zz$ coupling is smaller than the $yy$, 
which is what we have assumed already.

For a given value of $\Delta$, the line in the $(a,h)$ plane described by Eq.
(\ref{hd}) is called a disorder line because the direct product form of the 
ground state implies that the two-spin correlation function $<S_n^{\alpha} 
S_m^{\beta} > - <S_n^{\alpha} ><S_m^{\beta} >$ (with $\alpha , \beta =x,y,z$) 
is exactly zero if $m \ne n$. Hence the correlation length is extremely short.
The disorder line exists even for values of the spin larger than 1/2. 
Starting with the Hamiltonian in Eq. (\ref{ham3}), one finds a disorder line 
at the same value of h given in Eq. (\ref{hd}). The proof that it is 
a disorder line is similar to the proof given above for the spin-1/2 case if 
$\Delta \ge 0$. We will not study here how far the proof can be extended to
negative values of $\Delta$; for spin $S$, this requires an examination of the
spectrum of a two-site problem governed by a $(2S+1) \times (2S+1)$ 
dimensional Hamiltonian matrix.

\section{Ising transition}

If the magnetic field $h$ is increased even further, the system undergoes an 
Ising transition \cite{villain}. If the $yy$ and $zz$ couplings are equal 
(i.e., $1-a = \Delta$), this occurs at a saturation field $h_s = 2$, where 
there is transition to a state in which all the spins point along 
the $x$-axis. But if the
$yy$ and $zz$ couplings are not equal, there is no saturation of the spins for
any finite value of the field although the ground state expectation value of 
$S_n^x$ approaches 1/2 (as $(1-a-\Delta)^2 /h^2$) as $h$ goes to infinity. 
(This can be shown by considering a two-site system 
and doing perturbation theory in the limit $h \rightarrow \infty$.)
However, there is still a transition field $h_s$ beyond
which a $Z_2$ symmetry of a different kind is broken. To see this,
we consider a Hamiltonian ${\tilde H}$ which is dual to the Hamiltonian
given in Eq. (\ref{ham1}). This is given by
\bea
{\tilde H} ~&=&~ \sum_n ~[~ (1+a) ~T_n^x T_{n+2}^x ~+ ~\frac{1-a}{2} ~T_n^y
\non \\
& & ~~~~~~~~~ -~ 2 \Delta ~T_{n-1}^x T_n^y T_{n+1}^x ~- ~2 h ~T_n^x 
T_{n+1}^x ~]~.
\eea
This Hamiltonian is invariant under the global $Z_2$ transformation $T_n^x 
\rightarrow - T_n^x, T_n^y \rightarrow T_n^y, T_n^z \rightarrow - T_n^z$.
For $\Delta =0$, this $Z_2$ symmetry is known to be broken if $h$ is larger 
than a critical value $h_s$ \cite{villain}. We expect that this is will be
true even if $\Delta \ne 0$. The order parameter for this symmetry is 
\beq
m_x ~=~ [~ \lim_{n \rightarrow \infty} ~< T_0^x T_n^x > ~]^{1/2} ~.
\eeq

Note that in terms of the operators $S_n^x$, $T_0^x T_n^x$ is equal to 
a string of operators, $(1/4) \prod_{m=0}^{n-1} (2 S_m^x)$. Similarly, the
order parameter $(-1)^n S_0^y S_n^y$ in Eq. (\ref{order}) is equal to the
string of operators $((-1)^n /4) \prod_{m=1}^n (2 T_m^y)$.

\section{Discussion}

We have shown in this paper that the $XYZ$ spin-1/2 chain in a magnetic field
exhibits a gapless phase on a particular line. It would be interesting to use 
numerical techniques like the density-matrix renormalization group method 
\cite{pati} to examine various ground state properties of this model, in 
particular, to study the behavior of the order parameter defined in Eq. 
(\ref{order}), and to find out if there is indeed a Lifshitz point as 
conjectured in Fig. 2. 

Finally, the RG equations studied in this paper appear in other strongly
correlated systems, such as the problem of two spinless Tomonaga-Luttinger
chains with both one- and two-particle interchain hoppings \cite{nersesyan},
and one-dimensional conductors with spin-anisotropic electron interactions
\cite{giamarchi}. The gapless phase may therefore also appear in other systems.

\vskip .5 true cm
\centerline{\bf Acknowledgments}
\vskip .2 true cm

We would like to thank P. Fulde for hospitality in the Max-Planck-Institut
f\"ur Physik komplexer Systeme, Dresden, during the course of this work, and 
M. Barma and R. Narayanan for useful comments. A.D. acknowledges R. Oppermann
for interesting discussions, and Deutsche Forschungsgemeinschaft for financial 
support through project OP28/5-2. D.S. thanks the Department of Science and 
Technology, India for financial support through Grant No. SP/S2/M-11/00.

\end{document}